\documentstyle[preprint,aps,epsf]{revtex}

\tighten
\begin{document}
\draft
\title{\bf  Dynamical effects of a one-dimensional multibarrier
potential of finite range } 
\author{\bf  D.Bar$^{a}$ and L.P.Horwitz$^{a,b}$} 
\address{\bf $^a$Department of Physics, Bar Ilan University, Ramat Gan,
Israel \\ $^b$Raymond and Beverly Sackler Faculty of Exact Science, School of
Physics, Tel Aviv University, Ramat Aviv, Israel }
\maketitle 
\begin{abstract}   \noindent 
{\it   We discuss the properties  of a
large number $N$ of one-dimensional (bounded) locally periodic potential barriers in a 
finite interval. We show that 
the transmission coefficient, the scattering cross section 
$\sigma$,  and the resonances of  $\sigma$ depend sensitively upon the
ratio of the total spacing to the total barrier width. We also show that a 
 time dependent wave packet passing through the system of  potential barriers 
 rapidly spreads and deforms,  a criterion 
 suggested by Zaslavsky for chaotic behaviour. Computing the spectrum by 
 imposing (large) periodic boundary conditions we find a Wigner type
 distribution.       We investigate also the $S$-matrix
poles;  many resonances occur for certain values of the relative spacing between
the barriers in the potential. }
\end{abstract}
\noindent
\bigskip \noindent 
\protect \section{Introduction \label{sec1}}
Quantum systems with chaotic-like properties in the presence
of tunneling through a {\it single barrier} have been studied recently \cite{Schieve}.
The effect was attributed to the complexity of the wave function and its time
dependence in the neighbourhood of the barrier, where Zaslavsky's criterion for
the decrease of the Ehrenfest time \cite{Zasla} is expected to be satisfied. 
We discuss here several aspects of scattering from a one-dimensional locally
periodic potential 
barrier system \cite{Merzbacher,Cohen} which
is composed of a  large number $N$ of identical potential barriers densely
arrayed along a {\it finite} section of the $x$ axis.  We discuss both cases of 
finite and infinite $N$.   Parameters such as
the transmission coefficient, the scattering cross section and the energy
spectrum depend sensitively upon the ratio of the total spacing between the 
 potential
barriers to their total width. For example, it is shown in Section 2 for finite
$N$ and for both cases of $e>v$ and $v>e$ and in Sections 3, 4  and 5 
for  infinite $N$ that as this ratio  grows the transmission coefficient
tends  rapidly to unity.  An approaching particle can, therefore,  be transmitted 
unattenuated in
its amplitude through these  barriers without having to increase its
energy, even when $v>\!>e$ (for the $v>e$ case). Also,  it has been shown, using
the level statistics \cite{Reichl} of the energy spectrum of this dense system,  
that when this  ratio increases the dense system appears to 
become chaotic-like in the sense of ref \cite{Schieve}. These 
chaotic-like characteristics emerge also, as will be seen in Section 4,  when 
we study the passage of a Gaussian 
wave packet  through the dense system. 
We show also that  for both
cases  $e>v$ and $v>e$  the resonances  of the scattering cross section depend strongly 
 upon this
ratio. \par 
 Frishman and Gurvitz \cite{Frishman} have pointed out that the multiple 
barrier structure is important to study. They analyse the finite multiple 
well problem using a tight binding approximation, and find a miniband 
structure (with some similarity to the Kronig-Penney spectrum) which 
may correspond to the property of rapid approach to complete transmission 
that we find from certain values  of the ratio of spacing to barrier width
denoted in the following by $c$. 
The exact solution that we study permits us to investigate the very sensitive 
dependence of the transmission phenomena, as well as the distortion of wave 
packets with high precision for both small and large number N of barriers 
restricted to a fixed total interval. We find a simple form for the limit
$N \to \infty$, which retains the very sensitive dependence on $c$.     
 \par The configuration of a large density of one-dimensional barriers
in a small range can occur in physical systems with planar defects such as
successive evaporated layers or traps \cite{Smol,Bar1};  higher dimension 
analogs may also be realized. \par 
In section 2 we use a numerical model of this  system where the number $N$
of potential barriers is finite. We discuss both cases of $e>v$
and $v>e$. In Sections 3, 4   and 5 we discuss, using the transfer matrix method,
the limit of an infinite number  of potential barriers densely arrayed along
a  finite section of the $x$ axis. The $e>v$ case is discussed in Sections 3-4, 
and the
$v>e$ case in Section 5. In both cases we discuss the transmission
probability, the scattering cross section, the poles of this cross section 
 and the energy spectrum. We discuss
also the level statistics of the energy  spectrum of this dense system, and
the properties of a time dependent wave packet that passes through it. \par
We remark that although we study a system of $N$ barriers that is locally
periodic, all of these barriers are contained, even for $N \to \infty$, in a 
finite interval. The system is therefore not equivalent to a Kronig-Penney type
model \cite{Merzbacher};  the emergence of the band like structure as seen in Figure  
3 appears to
be due to the local (internal) periodicity but does not follow from global
crystal translation symmetry, this periodicity is, moreover, not in the particle
momentum, but in the total potential width. \par   We believe that similar results may also be obtained  for other similar
systems like, for example, the $N$ one dimensional periodic potential
wells in a finite interval. 
\protect \section{The $4N \times 4N$ matrix approach for the ensemble of potential
barriers \label{sec2}}
The array of $N$    potential barriers  discussed here  is  located 
 along  a finite section of the positive   $x$ 
axis beginning with the point $x=0$.  We denote 
the overall width of all the potential barriers by  $a$, and the total width 
of all the interim spaces separating them  by $b$. That is, in an 
array of $N$ potential barriers the width of each one is ${a\over N}$, and 
since in such an array there are $(N-1)$ separating spaces,  the width of each 
one is ${b\over (N-1)}$.  A sketch of our  array is shown   
in figure 1. To this array approaches from the negative half 
of the $x$ axis a plane wave   $e^{ikx}$, where 
 $k=({2me\over \hbar^2})^{{1\over 2}}$.  \par  
 We consider both cases: $e>v$, and 
 $v>e$, where $v$ is the constant height of each potential barrier, and 
 $e$ is the energy of the coming wave function.  We begin with the $e>v$  case and   
     write  the following 
 set of $4N$ simultaneous linear equations obtained from the boundary 
 conditions at the left and right hand sides  of all the $N$ potential barriers 
 \cite{Merzbacher,Cohen} (see figure 1). 
  
 \bigskip  \begin{eqnarray}
  1 + A &=& B + C     \nonumber \\
  ik - ikA &=& iqB -iqC    \nonumber \\
  Be^{i{qa\over N}}+Ce^{-i{qa\over N}} &=& De^{{ika\over N}}+Ee^{-{ika\over N}}   \nonumber \\
  iqBe^{i{qa\over N}}-iqCe^{-i{qa\over N}} &=& ikDe^{{ika\over
  N}}-ikEe^{-{ika\over N}}  \nonumber \\
  De^{ik({a\over N}+{b\over (N-1)})}+Ee^{-ik({a\over N}+{b\over (N-1)})} &=& 
  Fe^{iq({a\over N}+{b\over (N-1)})}+Ge^{-iq({a\over N}+{b\over (N-1)})}   \nonumber \\  
  ikDe^{ik({a\over N}+{b\over (N-1)})}-ikEe^{-ik({a\over N}+{b\over (N-1)})} &=& 
  iqFe^{iq({a\over N}+{b\over (N-1)})}-iqGe^{-iq({a\over N}+{b\over (N-1)})}   \nonumber  \\  
  ......................................&&.............................. \nonumber \\
 .....................................&&.................................\label{e1} \\
  ......................................&&.................................. \nonumber \\
  Re^{ik({a(N-1)\over N}+b)}+Se^{-ik({a(N-1)\over N}+b)}&=& Te^{iq({a(N-1)\over
  N}+b)}+
  Ue^{-iq({a(N-1)\over N}+b)}  \nonumber \\
  ikRe^{ik({a(N-1)\over N}+b)}-ikSe^{-ik({a(N-1)\over N}+b)} &=& 
  iqTe^{iq({a(N-1)\over N}+b)}-iqUe^{-iq({a(N-1)\over N}+b)}  \nonumber \\
  Te^{iq(a+b)}+Ue^{-iq(a+b)}&=& Ze^{ik(a+b)}  \nonumber \\
  iqTe^{iq(a+b)}-iqUe^{-iq(a+b)}&=& ikZe^{ik(a+b)},   \nonumber 
  \end{eqnarray}
  where $q=\sqrt{\frac{2m(e-v)}{\hbar^2}}$. 
  The former set can be written in a matrix form as  $Tx=\zeta$,  where 
  $T$ is the square matrix with $4N$ rows and $4N$ columnns 
   whose elements are  given 
  in the set (\ref{e1}). We denote by $x$  the unknown vector with the 
  $4N$ unknowns 
  $(A,B,C.....Z)$, and  
  $\zeta$ is the constant vector whose two first elements are $-1$ and $-ik$, and 
  all its other elements are zero.  As can be seen from the set (\ref{e1})  all the
  $4N$ 
  unknowns are obtained after dividing by the coefficient of the incoming 
  wave  so that $|A|^2$ and $|Z|^2$ (see 
  the first two and the last two equations of the set (\ref{e1})) are the coefficients 
  of reflection and transmission respectively.    In order to calculate 
  the value of $|Z|^2$ we have to solve the $4N$ simultaneous linear equations 
  $Tx=\zeta$. This can be done, especially for large $N$,   only by numerical 
  methods.   We denote  the constant 
  length  $a+b$  of the $x$ axis along which the system is
  arrayed by $L$, and define $b=ac$. 
  Thus,  we can express $a$ and $b$ in terms of
  $L$ and $c$ as follows \begin{equation} a=\frac{L}{1+c} \; \; \; \; \; 
  b=\frac{Lc}{1+c} \label{e2} \end{equation}  The continuous  curve in figure 
  2 shows the
  transmission coefficient $|Z|^2$ as a function of $c$ in the range 
  $1 \le c \le 35$  for $N=30$, and the
  dashed curve is for $N=40$.  The 
  other parameters are assigned the following values:  $v=100$, 
  $e=200$,  $\hbar=1$,  $m={1\over 2}$, $L=30$.  It is seen that the transmission
  coefficients,  for both values of $N$,  tend to unity when 
  $c$   grows, but for the larger $N$,  smaller values of $c$ suffice for 
    the transmission 
   coefficient  to approach unity.  That is,  {\it  when the 
      number of barriers increases, the
  approaching wave function passes unattenuated in its amplitude through these
  barriers even for relatively small values of $c$}. We note that we obtain the same result 
  of a unity value for the 
   transmission coefficient also when the restriction to a constant total
   length of the system is relaxed,  as seen from figure 3. The dashed curve in figure 3 shows 
   the transmission coefficient as a function of the total width $a$  when 
   $N=60$, and the continuous curve is for $N=120$. 
   For both curves we have assigned to $b$ the value  $b=\frac{a}{2}$. 
  The potential $v$ is 100  and   the energy $e$ is 200  as for figure 2. 
  From the dashed graph, for $N=60$, 
   we see that the transmission coefficient has an 
almost  constant 
  periodic pattern repeated as a function of $a$. In each one of these patterns 
  the  transmission coefficient oscillates near the value of 1, except at 
  the beginning 
  and end of each of these patterns where it drops  to zero. 
   In the continuous curve, for $N=120$, as in the dashed one, we 
  have also a 
 similar pattern repeated over the $a$ axis, but this time the width 
  of each such pattern is almost double.  Checking the  pattern of 
  the dashed 
  curve  we see that the 
    first  drop of the 
  transmission coefficient to zero occurs at $a \approx 9$, whereas the corresponding 
  drop in the continuous curve is at $a \approx 20$. 
  For $N=240$ (this graph is not shown here) the transmission coefficient 
  remains in the immediate
  neighbourhood of 1 when $1 \le a \le 40$. At about $a=40$ this 
  coefficient  drops to 0,  and then at $a=47$ it rises to 1 and remains in 
 the  neighbourhood of 1 until $a \approx 87$. Thus,  as $N$ 
 becomes
  larger the transmission coefficient appears to remain  in the neighbourhood of unity for 
  larger 
  intervals of $a$, so that as $N \to \infty$ these intervals would appear to 
  become infinite in
  extent.  In this limit one finds agreement with the closed form we
  obtain in  Eq (\ref{e15}). \par   
      We, now, discuss the  case $v>e$.  In this case the  set (\ref{e1}) has to be 
   changed to 
  take  into account the tunneling intervals, that is,  
  \bigskip \begin{eqnarray}     
  1 + A &=& B + C      \nonumber \\
  ik - ikA &=& -qB +qC     \nonumber \\
  Be^{-{qa\over N}}+Ce^{{qa\over N}} &=& De^{{ika\over N}}+Ee^{-{ika\over n}} \nonumber  \\
  -qBe^{-{qa\over N}}+qCe^{{qa\over N}} &=& ikDe^{{ika\over N}}-ikEe^{-{ika\over
  N}}  \nonumber \\
  De^{ik({a\over N}+{b\over (N-1)})}+Ee^{-ik({a\over N}+{b\over (N-1)})} &=& 
  Fe^{-q({a\over N}+{b\over (N-1)})}+Ge^{q({a\over N}+{b\over (N-1)})}   \nonumber \\  
  ikDe^{ik({a\over N}+{b\over (N-1)})}-ikEe^{-ik({a\over N}+{b\over (N-1)})} &=& 
  -qFe^{-q({a\over N}+{b\over (N-1)})}+qGe^{q({a\over N}+{b\over (N-1)})}  \nonumber   \\  
  ...................................&&................................... \nonumber \\
  ...................................&&......................................
  \label{e3}  \\
  ...................................&&...................................... \nonumber  \\
  Re^{ik({a(N-1)\over N}+b)}+Se^{-ik({a(N-1)\over N}+b)} &=& Te^{-q({a(N-1)\over
  N}+b)}+
  Ue^{q({a(N-1)\over N}+b)}  \nonumber \\
  ikRe^{ik({a(N-1)\over N}+b)}-ikSe^{-ik({a(n-1)\over n}+b)} &=& 
  -qTe^{-q({a(N-1)\over N}+b)}+qUe^{q({a(n-1)\over n}+b)} \nonumber \\
  Te^{-q(a+b)}+Ue^{q(a+b)} &=& Ze^{ik(a+b)}  \nonumber \\
  -qTe^{-q(a+b)}+qUe^{q(a+b)}&=& ikZe^{ik(a+b)} \nonumber    
  \end{eqnarray} 
 
In the set (\ref{e3}) $k$ is the same as the $k$ of the set (\ref{e1}) whereas $q$ is 
  $q=({2m(v-e)\over \hbar^2})^{1\over 2}$.   
  In  (\ref{e3})  the wave 
  functions  inside the potential barriers contain real exponentials. This changes 
  the  former 
  sinusoidal character of these wave functions (see the set (\ref{e1})) 
   to a hyperbolic one.  
 The continuous curve in figure  4   shows  the graph of the transmission 
 coefficient $|Z|^2$  as a
 function of $c$ for $N=30$, and the dashed curve is for $N=50$.
 The potential $v$ is 200, and the energy $e=180$. The total length of the
 system and the range of $c$ are the same as in figure 2, 
 that is, $L=30$, and   $1 \le c \le 35$.  
   We see  from both  graphs  that although $v>e$
 the  transmission coefficient tends to unity as $c$ grows, but
 this approach to unity is  faster  and for smaller 
 values of $c$,  when 
 $N$ is larger.  The
 same result is obtained if the condition of a constant total length of the
 system is relaxed as shown in figure 5,  which shows the transmission 
 coefficient as a function of the number $N$ of potential barriers of the
 system. In this figure we take the 
 total  width $a$ of all the potential barriers to be 8, and the total interval 
 $b$ to be $\frac{a}{2}$ ($c=\frac{1}{2})$.  The energy and the potential are assigned the values
 of 200 and 202 respectively.  In  this figure  we see that the 
  transmission coefficient  has  oscillating type behaviour  when the  
  number of  barriers is small.  At the larger values of $N$  the transmission 
  coefficient is in the 
close neighbourhood of unity.   \par  In summary, we see from this $4N \times 4N$ 
 matrix 
  method applied to both cases of $e>v$ and $v>e$, and for either a constant
  or variable total length of the system,  that when the ratio $c$ increases 
  the transmission coefficient tends to 1, and when the number of potential 
  barriers grows it tends to unity already at small values of $c$. 
\protect \section{The transfer matrix method for the $e>v$ case \label{sec3}}
We discuss, now, the multiple barrier system by the transfer matrix method 
\cite{Merzbacher,Cohen,Yu}, and in order to exploit its symmetry  the dense
array is assumed to be arranged between the points   
   $x=-\frac{a+b}{2}$ and   $x=\frac{a+b}{2}$, where $a$ and $b$ has the
same meaning as in the former section.   We discuss first the $e>v$ case.  
    Using the terminology of Merzbacher \cite{Merzbacher}  we can write  the 
following  transfer matrix equation which governs the behaviour of the bounded 
potential system.  \begin{equation} \label {e4} \left[ \begin{array}{c} A_{2n+1} \\ B_{2n+1} 
\end{array} \right] = P^{(n)}P^{(n-1)}\ldots P^{(2)}P^{(1)}
\left[ \begin{array}{c} A_0\\ B_0 \end{array} \right],  
\end{equation}  where $A_{2n+1}$ and $B_{2n+1}$ are the 
amplitudes of the transmitted and reflected parts respectively of the wave function 
from the $n$th potential barrier.   $A_0$ is the coefficient of the initial wave
that approaches the potential barrier system, and $B_0$ is the coefficient of the 
reflected wave from the first barrier.   $P^{(n)}$  is the product of three  two dimensional  matrices  
\begin{eqnarray}  
&& P^{(n)}= M_nT\grave M_n= \label{e5} \\ &&= \left[ \begin{array}{c c} e^{-ik(\frac{(n-1)b}{N-1}+
\frac{(2n-1)a}{2N})} & 0 \\ 0 & e^{ik(\frac{(n-1)b}{N-1}+
\frac{(2n-1)a}{2N})} \end{array} \right]\left[ \begin{array}{c c}T_{11}&T_{12}
\\T_{21}&T_{22}\end{array} \right]\left[ \begin{array}{c c}
e^{ik(\frac{(n-1)b}{N-1}+
\frac{(2n-3)a}{2N})} & 0 \\ 0 & e^{-ik(\frac{(n-1)b}{N-1}+
\frac{(2n-3)a}{2N})} \end{array} \right] \nonumber \end{eqnarray}
 The middle matrix $T$ does not depend on $n$ \cite{Remark1} and its components are given by 
    \begin{eqnarray}  &&T_{11} = 
 \cos(\frac
{aq}{N})+i\frac{\xi}{2}\sin(\frac{aq}{N}), \; \; \;   T_{12} =i\frac{\eta}{2}
\sin(\frac{aq}{N}) \label{e6} \\  &&T_{21} = -i\frac{\eta}{2}
\sin(\frac{aq}{N}), \; \; \;  T_{22} = 
 \cos(\frac
{aq}{N})-i\frac{\xi}{2}\sin(\frac{aq}{N}) \nonumber  \end{eqnarray}
Here   $k$ is
  $\sqrt{\frac{2me}{\hbar^2}}$, 
  $q$ is $\sqrt{\frac{2m(e-v)}{\hbar^2}}$,    
 and $\xi$ and  $\eta$ are given by \begin{equation} \label{e7} \xi=\frac{q}{k}+\frac{k}{q}, \quad 
  \eta=\frac{q}{k}-\frac{k}{q} \end{equation}  
As can be seen from Eq (\ref{e5})  the product of each  neighbouring diagonal matrices 
$\grave M_nM_{n-1}$ is constant for each $n$. That is, 
$\grave M_nM_{n-1}= 
    \left[
\begin{array}{c c}e^{\frac{ikb}{N-1}}&0\\
0&e^{-\frac{ikb}{N-1}} \end{array} \right]$. 
 Thus,   we  may 
write Equation (\ref{e4}) as \begin{eqnarray}  &&\left[
\begin{array}{c} A_{2n+1} \\ B_{2n+1} 
\end{array} \right]=\left[ \begin{array}{c c} e^{-ik(a+b-\frac{a}{2N})}
 & 0 \\0&e^{ik(a+b-\frac{a}{2N})} \end{array} \right] \left[ \begin{array}{c c}
T_{11}&T_{12}
\\T_{21}&T_{22}\end{array} \right] \Biggl( \left[ \begin{array}{c c}
e^{\frac{ikb}{N-1}}&0\\
0&e^{-\frac{ikb}{N-1}} \end{array} \right] \left[ \begin{array}{c c}
T_{11}&T_{12}
\\T_{21}&T_{22}\end{array} \right] \Biggr) ^{n-1} \cdot \nonumber  \\ 
&& \cdot \left[ \begin{array}{c c}
e^{\frac{-ika}{2N}}&0\\
0&e^{\frac{ika}{2N}} \end{array} \right]\left[ \begin{array}{c} A_0 \\ B_0 
\end{array} \right]  \label{e8} 
 \end{eqnarray} 
   If we take the limit of a very
 large $N$,  we obtain for the right hand side of the
 potential barrier system at the point $x=\frac{a+b}{2}$ where  $n=N$
\begin{equation}  \left[
\begin{array}{c} A_{2N+1} \\ B_{2N+1} 
\end{array} \right]=\left[ \begin{array}{c c} e^{-ik(a+b)}
 & 0 \\0& e^{ik(a+b)} \end{array} \right] \Biggl( \left[ \begin{array}{c c}
e^{\frac{ikb}{N}}&0\\
0&e^{-\frac{ikb}{N}} \end{array} \right] \left[ \begin{array}{c c}
T_{11}&T_{12}
\\T_{21}&T_{22}\end{array} \right] \Biggr) ^{N} \left[ \begin{array}{c} A_0 \\ B_0 
\end{array} \right]  \label{e9} 
 \end{equation} 
 We note that the last equation can be discussed from the eigenvalue point of
 view  \cite{Merzbacher,Cohen}.  That is, by finding the appropriate eigenvalues
 from the suitable characteristic equation (a similar method has been applied to the 
  finite $N$ potential barrier  system in \cite{Cohen}). In the following we
  adopt a more analytical and exact approach that yields the same results
  obtained from the former numerically-oriented method. \par 
The expression under the exponent $N$ in Eq (\ref{e9}) can be written, using the
set (\ref{e6}), in the limit of very large $N$,  as 
\begin{eqnarray}   
&& \Biggl( \left[ \begin{array}{c c}
e^{\frac{ikb}{N}}&0\\
0&e^{-\frac{ikb}{N}} \end{array} \right] \left[ \begin{array}{c c}  
 \cos(\frac
{aq}{N})+i\frac{\xi}{2}\sin(\frac{aq}{N}) &  i\frac{\eta}{2}
\sin(\frac{aq}{N}) \\ -i\frac{\eta}{2}
\sin(\frac{aq}{N})& \cos(\frac{aq}{N})-i\frac{\xi}{2}\sin(\frac{aq}{N}) 
\end{array} \right] \Biggr) ^{N} \cong \nonumber \\ && 
\cong \Biggl( \left[ \begin{array}{c c}
(1+\frac{ikb}{N})(1+\frac{i\xi}{2}(\frac{aq}{N}))
&i(1+\frac{ikb}{N})\frac{\eta}{2}(\frac{aq}{N}) \\ 
-i(1-\frac{ikb}{N})\frac{\eta}{2}(\frac{aq}{N}) &
(1-\frac{ikb}{N})(1-\frac{i\xi}{2}(\frac{aq}{N})) 
\end{array} \right] \Biggr) ^{N} \cong \label{e10} \\&& \cong 
\Biggl( 1+\frac{i}{N} \Biggl( \left[ \begin{array}{c c} kb+aq\frac{\xi}{2}&
aq\frac{\eta}{2} \\ -aq\frac{\eta}{2}& -(kb+aq\frac{\xi}{2}) 
\end{array} \right] \Biggr) \Biggr)^N
\cong (1+\frac{i}{N}((kb+aq\frac{\xi}{2})\sigma_3+iaq\frac{\eta}{2}\sigma_2))^N,
\nonumber  
\end{eqnarray}
where  $\sigma_2$ and 
$\sigma_3$ are the standard  Pauli matrices $ \sigma_2=
\left[ \begin{array}{c c} 0&-i\\i&0 \end{array} \right]$,  
$\sigma_3=\left[ \begin{array}{c c} 1&0\\0&-1 \end{array} \right] $.  Using the 
relation $\lim_{n\to\infty}(1+\frac{c}{n})^n=e^c$, where $c$ is some (possibly 
matrix-valued) constant 
we obtain from equations (\ref{e9}),(\ref{e10}) 
\begin{equation} \left[ \begin{array}{c} A_{2N+1} \\ B_{2N+1} 
\end{array} \right] =\exp(-ik(a+b)\sigma_3)\exp(i((kb+\frac{aq\xi}{2})\sigma_3+
\frac{iaq\eta}{2}\sigma_2)) \left[ \begin{array}{c} A_{0} \\ B_{0}
\end{array} \right] \label{e11} \end{equation}
We denote the two coefficients in   the second exponent as 
\begin{equation} \label{e12} f=kb+aq\frac{\xi}{2},  \; \; \; d=aq\frac{\eta}{2};
\end{equation}  making use of the relation \begin{equation} 
\label{e13} (f\sigma_3+id\sigma_2)^2=f^2-d^2=\phi^2, \end{equation} 
we can expand the second exponent on the right hand side of
Eq (\ref{e11}) in a Taylor series.   After collecting 
corresponding terms we obtain 
\begin{equation} \label{e14} e^{i((kb+aq\frac{\xi}{2})\sigma_3+
iaq\frac{\eta}{2}\sigma_2)}=\cos(\sqrt{f^2-d^2})+\frac{i(f\sigma_3+id\sigma_2)}
{\sqrt{f^2-d^2}}\sin(\sqrt{f^2-d^2}) \end{equation} 
  Defining $z=k(a+b)$,   we , therefore,  obtain   
\begin{equation} \label{e15} \left[ \begin{array}{c} A_{2N+1} \\ B_{2N+1}
\end{array} \right]=\left[ \begin{array}{c c}
e^{-iz}(\cos{\phi}+if\frac{\sin(\phi)}{\phi}) &ie^{-iz}d\frac{\sin(\phi)}{\phi} 
\\ -ie^{iz}d\frac{\sin(\phi)}{\phi}&e^{iz}(\cos{\phi}-if\frac{\sin(\phi)}{\phi})
\end{array} \right]\left[ \begin{array}{c} A_{0} \\ B_{0}
\end{array} \right] \end{equation} 
The corresponding expression for a single barrier of the same total width and
location is \cite{Merzbacher} \begin{equation} \label{e16} \left[ \begin{array}{c}  A \\  B 
\end{array} \right] = \left[ \begin{array}{c c}
e^{-iz}(\cos(aq)+i\frac{\xi}{2}\sin(aq)) &ie^{-iz}\frac{\eta}{2}\sin(aq) 
\\ -ie^{iz}\frac{\eta}{2}\sin(aq)&e^{iz}(\cos(aq)-i\frac{\xi}{2}\sin(aq))
\end{array} \right]\left[ \begin{array}{c} A_{0} \\ B_{0}
\end{array} \right] \end{equation} One sees that the internal structure of the multiple
barrier, in the limit of $N\to \infty$, is  different (they coincide only if
$b = 0$).
The determinant of the matrix on the right
hand side of   Eq (\ref{e15})  is unity.  \par
 As we have seen,   the equations
(\ref{e9})-(\ref{e15}) were concerned with expressing the amplitudes of the
transmitted and reflected parts $A_{2N+1}$, $B_{2N+1}$ of the wave function at
the right hand side of the bounded system (at the point $\frac{(a+b)}{2}$) as  
functions
of $A_0$, and $B_0$ at the left hand side of this system (at the point 
$-\frac{(a+b)}{2}$).  We can find these amplitudes at an arbitrary point
$-\frac{a+b}{2}<x<\frac{a+b}{2}$ by using the property of the system that 
it is an infinite sequence of 
potential barriers  bounded at  two sides, so the point $x$ is
associated with some  barrier $n$ and may be written as 
 $x=\pm n(\frac{a}{N}+\frac{b}{N-1})=\pm np$, where
 $p=\frac{a}{N}+\frac{b}{N-1}$,    
 and $n$ is in the range $1 \le n \le \frac{N}{2}$. The potential cycle  $p$ 
can be expressed in terms of the total length $L$ 
as $p=\frac{L}{N}$, so    
$x=\pm n\frac{L}{N}$,    or   $n=\pm \frac{xN}{L}$.   
   Since $x$ and $L$ are finite numbers $n$ must be  infinite
 if $N$ is. Thus,  the amplitudes
of the transmitted and reflected parts  $A_{2n+1}$, $B_{2n+1}$ of the wave
function at the $n$-th potential barrier can be  written as  (compare with Eq 
(\ref{e11}))  
\begin{eqnarray} 
&& \left[
\begin{array}{c} A_{2n+1} \\ B_{2n+1} 
\end{array} \right]=\left[ \begin{array}{c c} e^{-ik(a+b)}
 & 0 \\0& e^{ik(a+b)} \end{array} \right]
 (1+\frac{i}{N}((kb+aq\frac{\xi}{2})\sigma_3+iaq\frac{\eta}{2}\sigma_2))^n  
 \left[ \begin{array}{c} A_0 \\ B_0 
\end{array} \right]=  \nonumber  \\ && =\exp(-ik(a+b)\sigma_3)(1+
\frac{i}{N}((kb+aq\frac{\xi}{2})\sigma_3+iaq\frac{\eta}{2}\sigma_2))^{\pm \frac{xN}{L}}
\left[ \begin{array}{c} A_0 \\ B_0 \end{array} \right]= \label{e17} \\ && =
\exp(-ik(a+b)\sigma_3)\exp(\pm \frac{ix}{L}
(f\sigma_3+id\sigma_2))\left[ \begin{array}{c} A_0 \\ B_0 \end{array} \right] \nonumber
\end{eqnarray}  We have used Eqs (\ref{e12}) and 
the relation 
$\lim_{n\to\infty}(1+\frac{c_1}{n})^{nc_2}=e^{c_1c_2}$, where $c_1$ and $c_2$
are arbitrary finite (possibly matrix-valued) constants. Now if we define $f_1=\frac{fx}{L}$, 
$d_1=\frac{dx}{L}$ we obtain $f_1^2-d_1^2=\frac{x^2}{L^2}\phi^2=\phi_1^2$,
where $\phi$ is given by Eq (\ref{e13}). Thus, we may use all the equations
written before (for the right hand side of the dense system   at the point 
$x=\frac{a+b}{2}$)  also for an arbitrary point 
$-\frac{a+b}{2}<x<\frac{a+b}{2}$.   This result provides a closed form for the
wave function in the potential region which we shall discuss further in a 
succeeding publication. \par 
Now,  defining  
$e^{i\kappa}=\frac{(\cos(\phi)+if\frac{\sin(\phi)}{\phi})}
{\sqrt{(\cos^2(\phi)+f^2\frac{\sin^2(\phi)}{\phi^2})}}$, 
  we can find from Eq (\ref{e15}) the transmission probability at the point 
$x=\frac{a+b}{2}$   by noting that  at this point we have zero 
reflection, so  $B_{2N+1}=0$. Thus, we
obtain for this probability
\begin{equation} \label{e18} |\frac{A_{2N+1}}{A_0}|^2=|\frac{1}{e^{iz}
(\cos(\phi)-
if\frac{\sin(\phi)}{\phi})}|^2= |\frac{e^{i(\kappa-z)}}{\sqrt{\cos^2(\phi)+
f^2\frac{\sin^2(\phi)}{\phi^2}}}|^2 =\frac{1}{1+\frac{d^2\sin^2(\phi)}
{\phi^2}},  \end{equation} where we have  used Eqs (\ref{e12}). 
We  see that the  transmission
probability reduces,  when $b=0$, 
to the known transmisson probability  of the one potential barrier system which
is located at the same place   and exposed to the same wave function as the infinite
potential barrier system
\cite{Merzbacher,Cohen},   \begin{equation} \label{e19}  |\frac{A}{A_0}|^2=
\frac{1}{\cos^2(aq)+\frac{\xi^2\sin^2(aq)}{4}}.  \end{equation}  We will see that the presence
of a finite $b$ in Eq (\ref{e18}) results in a new possibility for the 
transmisson probability to reach unity  without having to increase the energy as in the
one potential barrier system. 
  Clearly, if $a \to 0$ (no potential barrier),  
 the transmission
coefficient goes trivially to 1. 
Figure 6 
shows a three dimensional graph of the transmission probability from Eq
(\ref{e18}) as a function of the energy $e$ and $c$. The total length is $L=70$, 
$v=60$, the range of $e$ is $61 \le e \le 120$, and that of $c$ is  
$0.01 \le c \le 5$. As expected the   transmission probability tends to unity 
when 
the energy $e$ grows, but as seen from the graph it tends faster, even
critically,  to the
neighbourhood of  unity (even
for small $e$) as $c$ increases through relatively small values (i.e., this
effect is not simply due to $a \to 0$), that is, as the total spacing becomes larger 
(see Eq (\ref{e2})).  This is in agreement with  the results obtained in the previous 
section for the $e>v$ case (see figures 2-3). 
 \par 
\protect \section{Scattering cross section \label{sec2}} 
 
We  now study the scattering cross section  of the bounded potential barrier system. For
this we use the $S$ matrix which connects the outgoing waves $A_{2N+1}$,
 and $B_0$ at the two sides of the system to the ingoing ones $B_{2N+1}$ and
 $A_0$. That is,   \begin{equation}  \label{e20} \left[
\begin{array}{c} A_{2N+1} \\ B_{0} 
\end{array} \right]=\left[ \begin{array}{c c} S_{11}&S_{12} \\ S_{21} &S_{22} 
\end{array} \right]\left[ \begin{array}{c} A_{0} \\ B_{2N+1} 
\end{array} \right] \end{equation} 
Using the last equation together with equation (\ref{e15}) (we  denote the
 two dimensional matrix from Eq (\ref{e15}) by $Q$ where $\det Q=1$)  one 
 obtains for the
four components of the matrix $S$ \cite{Merzbacher} \begin{equation}  S_{11}=Q_{11}-
\frac{Q_{12}Q_{21}}{Q_{22}}=\frac{1}{Q_{22}}, \; \; \; \; 
S_{12}=\frac{Q_{12}}{Q_{22}}, \; \; \; \; S_{21}=-\frac{Q_{21}}{Q_{22}}, \;
\; \; \; S_{22}=\frac{1}{Q_{22}} \label{e21} \end{equation}   Now, in order to 
find the phase shifts we have to find
the eigenvalues $\lambda_{\pm}$ from the following equation 
 $\det(S-\lambda I)=\det\left[ \begin{array}{c c} 
S_{11}-\lambda&S_{12} \\ S_{21} &S_{22}-\lambda 
\end{array} \right] =0 $.   That is,   \begin{eqnarray} && \lambda_{\pm}=\frac{1}{Q_{22}}(1\pm
i\sqrt{|Q_{12}|^2})=\frac{1 \pm \frac{id\sin(\phi)}{\phi}}{e^{iz}(\cos(\phi)+
if\frac{\sin(\phi)}{\phi})}=\frac{(1\pm \frac{id\sin(\phi)}{\phi})
e^{-iz}(\cos(\phi)-\frac{if\sin(\phi)}{\phi})}{\cos^2(\phi)+
\frac{f^2\sin^2(\phi)}{\phi^2}}= \nonumber \\ &&=
\frac{\cos(z)\cos(\phi)+\frac{f\sin(z)\sin(\phi)}{\phi} \pm 
d(\frac{\sin(z)\sin(\phi)\cos(\phi)}{\phi}-
\frac{f\cos(z)\sin^2(\phi)}{\phi^2})}{\cos^2(\phi)+
\frac{f^2\sin^2(\phi)}{\phi^2}}
+\label{e22} \\&&+\frac{i(\pm d(\frac{\sin(\phi)\cos(z)\cos(\phi)}{\phi}+ 
\frac{f\sin(z)\sin^2(\phi)}{\phi^2})-\sin(z)\cos(\phi)+
\frac{f\cos(z)\sin(\phi)}{\phi})}{ \cos^2(\phi)+
\frac{f^2\sin^2(\phi)}{\phi^2}} \nonumber 
 \end{eqnarray} 
According to the conventional phase shift theory  \cite{Merzbacher,Cohen}  
 $\lambda_{\pm} =e^{2i\delta_{\pm}}$,  
     where $\delta_{\pm}$ are the phase shifts that correspond to the 
eigenvalues $\lambda_{\pm}$. From the last  
relations we obtain \begin{equation} \label{e23} \cos(2\delta_{\pm})=
\frac{\cos(z)\cos(\phi)+\frac{f\sin(z)\sin(\phi)}{\phi} \pm 
d(\frac{\sin(z)\sin(\phi)\cos(\phi)}{\phi}-
\frac{f\cos(z)\sin^2(\phi)}{\phi^2})}{\cos^2(\phi)+
\frac{f^2\sin^2(\phi)}{\phi^2}} \end{equation}
 \begin{equation} \label{e24} \sin(2\delta_{\pm})=
\frac{\pm d(\frac{\sin(\phi)\cos(z)\cos(\phi)}{\phi}+ 
\frac{f\sin(z)\sin^2(\phi)}{\phi^2})-\sin(z)\cos(\phi)+
\frac{f\cos(z)\sin(\phi)}{\phi}}{ \cos^2(\phi)+
\frac{f^2\sin^2(\phi)}{\phi^2}}  
 \end{equation} 
     The scattering amplitude is given by   
  $S-1=e^{2i\delta_{\pm}}-1=2ie^{i\delta_{\pm}}\sin(\delta_{\pm})=
 2\pi iT_{\pm}$, and 
 the cross section $\sigma_{\pm}$ is then obtained as \begin{equation} \label{e25} 
 \sigma_{\pm}=4\pi^2|T_{\pm}|^2=4\sin^2(\delta_{\pm})=2(1-\cos(2\delta_{\pm}))  
 \end{equation} 
 We see  that $\sigma_{\pm} \to 0$ for $e \to \infty$, 
since $d \approx O(\frac{1}{\sqrt{e}}) \to 0$ and $\frac{f}{\phi} \to 1$. It
is further clear that for large $e$ the period of oscillation grows;  the
oscillations with respect to $e$ go as $\sqrt{e}$,  and the period, for which 
$\sqrt{e+\Delta}=\sqrt{e}+2\pi$, is determined by $\Delta=4\pi^2+4\pi\sqrt{e}$, 
 therefore grows.  
 Figure 7 shows a graph of $\sigma_+$ as a function of the energy $e$ (the same
graph is obtained also for $\sigma_-$). 
 The total 
length $L$ is 70, and $a=40$, so $b=30$. The potential $v$ is taken to be  70, 
and the range of $e$ is  
$71 \le e \le 1000$. One sees the increase in period on this graph, but the
decrease in amplitude would not become visible until $e >\!> a^2v^2/4$,  i.e; 
for our case $e >\!> 2\cdot 10^6$.  
 An interesting property of $\sigma_{\pm}$ emerges when we relax  the constraint of  
constant  $L=a+b$. In this case we find that  the dependence of the cross 
 section   upon $b$  is  different from that upon $a$. That is, for the 
 same value of
 $a$ the cross sections $\sigma_{\pm}$, as  functions of the energy $e$,  depend  
also  on $b$ only in a finite 
   specific range that depends upon the value of $a$. For example, for $v=70$ 
   and $a=10$, 
   the
   cross section $\sigma_+$ changes with $b$  in the  range of $0 \le b
   \le 3.5a$,  and  for $b > 3.5a$ the change in $\sigma_+$ is so small that, 
    as a function of $e$,  can be considered  constant.  
   The same thing
   can be said also for $\sigma_-$.  
   We find that    
     the cross sections  
 depend on the total width $a$ (for $b$ fixed)  in such a manner that the  
   periods  of $\sigma_{\pm}$ are inversely proportional to $a$.  
 That is, as the total width $a$ of the potential barrier 
 system  grows the growth rate (with $e$) of the period of $\sigma_{\pm}$
  becomes smaller.
  \par   We discuss now the energy level statistics \cite{Reichl} of the bounded
dense array. To study this problem  we use the $S$-matrix and the boundary value
conditions at two remote boundaries  of the system. That is, using periodic boundary conditions 
at the points $|x|=C$, 
where $C$ is much
 larger than the size $L=a+b$ of the system, 
we obtain 
 $ A_{2N+1}e^{ikC}=A_0e^{-ikC},  \; \; \; \; 
B_{2N+1}e^{-ikC}=B_0e^{ikC} $.  
Thus, using the last two relations,  and expressing the
components of $S$ in terms of those of $Q$ (see  
Eq (\ref{e21})),  
 we write Eq 
(\ref{e20}) as  \begin{equation}  \label{e26} \left[
\begin{array}{c} A_{2N+1} \\ B_{0} 
\end{array} \right]=e^{2ikC} \left[ \begin{array}{c c} S_{11}&S_{12} \\ S_{21} &S_{22} 
\end{array} \right]\left[ \begin{array}{c} A_{2N+1} \\ B_{0} \end{array} \right]=
\frac{e^{2ikC}}{Q_{22}}\left[ \begin{array}{c c} 1&Q_{12} \\ -Q_{21} &1 
\end{array} \right]\left[ \begin{array}{c} A_{2N+1} \\ B_{0}
\end{array} \right] \end{equation} 
To obtain a non trivial solution for the vector $\left[
\begin{array}{c} A_{2N+1} \\ B_{0} 
\end{array} \right]$  we have to solve the following equation;  
$ \det \left[ \begin{array}{c c} 
\frac{e^{2ikC}}{Q_{22}}-1& \frac{e^{2ikC}Q_{12}}{Q_{22}}\\
-\frac{e^{2ikC}Q_{21}}{Q_{22}}  & \frac{e^{2ikC}}{Q_{22}}-1
\end{array} \right] =0 $. The last equations, after  
 substituting for the $Q$'s from Eq (\ref{e15}),  
becomes
\begin{eqnarray} && \det\left[ \begin{array}{c c} 
\frac{e^{2ikC}}{Q_{22}}-1& \frac{e^{2ikC}Q_{12}}{Q_{22}}\\
-\frac{e^{2ikC}Q_{21}}{Q_{22}}  & \frac{e^{2ikC}}{Q_{22}}-1
\end{array} \right]=\frac{e^{4ikC}}{Q^2_{22}}-\frac{2e^{2ikC}}{Q_{22}}+1+
\frac{Q_{12}Q_{21}e^{4ikC}}{Q^2_{22}}=\label{e27} \\ 
&&=\cos(4kC)(1+\frac{d^2\sin^2(\phi)}{\phi^2})+\cos(2z)(\cos^2(\phi)-
\frac{f^2\sin^2(\phi)}{\phi^2})+\frac{2f\sin(\phi)}{\phi}(\sin(2z)\cos(\phi)-
\nonumber \\ && -\sin(2kC+z))-2\cos(2kC+z)\cos(\phi)+ 
i(\sin(4kC)(1+\frac{d^2\sin^2(\phi)}{\phi^2})+\sin(2z)(\cos^2(\phi)-
\nonumber \\ &&-
\frac{f^2\sin^2(\phi)}{\phi^2}) +\frac{2f\sin(\phi)}{\phi}
(\cos(2kC+z)-\cos(2z)\cos(\phi))-2\sin(2kC+z)\cos(\phi)) =0 \nonumber 
\end{eqnarray}
In order to obtain the spectrum 
we solve,
numerically, 
the last equation for the energies that satisfy both its real and imaginary
parts.    Obtaining  these energies we use the 
unfolding procedure \cite{Reichl}
to transform to the more appropriate energies from which we  may obtain energy 
level distribution. We find that  the distribution of the energy 
level spacings
depends sensitively upon the value of $c$. For small $c$ (small total interval
$b$ and large total width $a$) the relevant distribution is more of the  
Poisson distribution type \cite{Reichl}  than of the chaotic Wigner one, 
whereas when $c$ increases the corresponding distribution is more of the 
Wigner
type than of the Poisson one. Figure 8 shows a histogram form of the level spacings 
distribution \cite{Reichl} of 102 energy levels  obtained  numerically 
 for both cases of $e>v$ and $v>e$. The potential height is here taken to be 
 $v=120$, and 
$C=90$, $L=20$  $c=19$. The continuous curve is the chaotic Wigner
distribution \cite{Reichl} as obtained from a random matrix model \cite{Dyson},
and the dashed one is the Poisson distribution \cite{Reichl}. One can see that the 
histogram-form curve
resembles the chaotic Wigner one.  The strong peak at 0.5 appears not consistent
 with the Poisson (dashed) curve.  We remark that 
including levels only
for $e>v$ results in a distribution which is not clearly of Wigner type; if 
we select only  $v>e$ the distribution appears more clearly of Wigner type.   
  We  note that when we use Eq (\ref{e27})  
for the $v>e$ case, the $f$, $d$ and $\phi$ that must be substituted in this
equation are not those defined by equations (\ref{e12}),(\ref{e13}), but those
defined in the following section in Eq  (\ref{e34}).     
\par   
In order to investigate further the properties of the bounded dense system we
  have passed a Gaussian wave packet through it and study  its behaviour in the
  bounded potential region. We have used the complex packet \begin{equation}
  \label{e28}
  \phi(x,t,x_0,p_0,w_0)=\frac{\sqrt{w_0}\pi^{\frac{1}{4}}e^{-\frac{p_0^2}{4w_0^2}}
 e^{\frac{w_0^2(i(x_0-x)-\frac{p_0}{2w_0^2})^2}{1-2itw_0^2}}}{\sqrt{1-2itw_0^2}},
\end{equation} where $x_0$ is the initial mean position of the packet in coordinate
 space, and $p_0$ and $w_0$ are the initial mean momentum, and  
 initial width (uncertainty) of the momentum respectively in $p$ space. 
  For our numerical simulations we have discretized space and time
 with a resolution of $dx=\frac{1}{7}$ and $dt=\frac{1}{50}$. This resolution
 ensures the condition $dt<dx^2$ which is necessary for a stable and steady
 performance of the numerical method used here \cite{Maple}. For the other 
 parameters we choose  $m=\frac{1}{2}$, $w_0=\frac{1}{2}$, $x_0=-10$, $c=2.333$ 
 $v=2$, and
$p_0=3$. The last two chosen values ensure the condition of $e>v$. The dense
system  is arrayed between the points $x=-10$, and  $x=10$. The units we are
using for length and time are therefore; $x=\frac{x_{cm}}{\hbar}$ and 
$t=\frac{t_{sec}}{m\hbar}$ (we take $p$ to be momentum in units $[mv]$ and $w_0$
the dispersion in $p$).  With this scale, we see that velocities in 
$\frac{cm}{sec}$ are related to our parametric velocities by 
$\frac{\Delta x_{cm}}{\Delta t_{sec}}=\frac{1}{m}\frac{\Delta x_{cm}}{\Delta
t_{sec}}$.  
During and after 
the passage of  the wave packet  through the potential region
 its
initial Gaussian form is strongly  deformed.   The point-type curve in figure 9
 shows the form of the density of the wave packet which evolves from Eq (\ref{e28}) when 
 the number of
 potential barriers is $N=4$,  and  the continuous graph is the form of this 
 density  for $N=150$. The potential barriers arrayed between $x=-10$ and
 $x=+10$ are not shown. Both curves in figure 9 are for the same time of
 $t=5.8$. Comparing these curves we see that for
large $N$  the wave packet expands across  the whole 
potential region   more rapidly than  the expansion for   $N=4$.   It has been
suggested by Zaslavsky \cite{Zasla} (see also discussion in Ref \cite{Schieve}) 
that this behaviour is characteristic of 
a system with classical Hamiltonian of chaotic   type.    
 Moreover, we see that the structure of the wave function for $N=150$ is of a much
 higher degree of complexity. For larger values of $t$ (not shown) one sees that
 the forms of the transmitted and reflected waves are also of higher 
 complexity. These results 
 depend sensitively on the value of $c$. 
The effect is most pronounced when $c$ is in the neighbourhood of 4,  
 whereas, as $c$ grows this effect  diminishes until it
completely disappears for very large $c$.   Moreover, we obtain a significant 
transmission of the wave packet through the potential region even for the case 
of $v>e$, and the part of the wave packet that passes through this region 
increases as
the number $N$ or $c$ or both of them grow. 
Note that we have found that the transmission of  a 
 plane wave grows rapidly with $c$,  whereas it may be  poorly
 transmitted  when $c$ is very small;    in accordance  with the wave
 packet behavior just described.   \par
We now study  the problem of resonances associated with the dense system.  
In order to find them we find the resonances  of the cross section 
 $\sigma_{\pm}$ (see Eq
 (\ref{e25})).   Using Eqs (\ref{e12}),(\ref{e13}) and 
 Eq (\ref{e23}) (or Eq (\ref{e24}))  
 we find that these
  are  found
 at the values of the energies $e$ that satisfy the following equation 
 \begin{equation} 1+\frac{d^2\sin^2(\phi)}{\phi^2}=0 \label{e29} \end{equation} 
    Eliminating the total interval $b$  (see Eq (\ref{e2})) from the following 
 equations and  substituting in 
 Eq (\ref{e29}) for $d$ and $\phi$ from Eqs (\ref{e12})-(\ref{e13}),  
 and also  for    $q$ and $k$   
   we obtain  
 \begin{eqnarray}  &&\sin(\sqrt{\frac{(a+ac+cL)}{(1+c)^2}
 (e(a+ac+cL)-av(1+c))}) = \label{e30} \\ &&
= \pm \frac{2i}{av(1+c)}\sqrt{e(a+ac+cL)(e(a+ac+cL)-va(1+c))} \nonumber 
 \end{eqnarray}
 The last equation can be, of course,  valid only if the energy $e$ is complex.  
  We denote this energy as $e=e_1+ie_2$. 
The components $e_1$ and $e_2$ are found  (see Appendix A) by solving the following two 
simultaneous equations.
\begin{equation} \label{e31} 
 \sin(r_1^{\frac{1}{2}}
  \cos(\frac{\phi_1+2\pi k}{2}))\cosh(r_1^{\frac{1}{2}}
  \sin(\frac{\phi_1+2\pi k}{2}))=\pm r^{\frac{1}{2}}_2\cos(\frac{\phi_2+2\pi k}{2})
  \end{equation} \begin{equation}  \label{e32} \cos(r_1^{\frac{1}{2}}
  \cos(\frac{\phi_1+2\pi k}{2}))\sinh(r_1^{\frac{1}{2}}
  \sin(\frac{\phi_1+2\pi k}{2}))=\pm r^{\frac{1}{2}}_2\sin(\frac{\phi_2+2\pi k}{2}), 
  \end{equation} 
   where $k=0, 1$ and $r_1$,  $\cos (\phi_1)$,  $\sin (\phi_1)$,  $r_2$,  $\cos (\phi_2)$, and 
 $\sin (\phi_2)$ are given respectively by Eqs (\ref{eA3})-(\ref{eA8}) in  Appendix A.  
     We find  numerically  that 
    there is no solution to equations (\ref{e31}),(\ref{e32}) for 
    very large values
of $c$.  
 The allowed range of $c$,   for which these equations  
 may be satisfied,  depends upon the value of the total
length of the system $L$; 
as $L$ increases the allowed range of $c$ 
expands. 
For all other values of  $c$ outside these ranges  
 we  find no pole  that satisfy the simultaneous equations (\ref{e31}),(\ref{e32}). 
 We note that the poles are more frequent at the middle sections of these ranges
 than at their ends.     As noted  the absolute values of the complex energy 
  must be greater than $v$ since we deal
 here with the $|e|>v$ case.  It can be shown (see Appendix B) that as long as 
 these absolute values  are not
 very much larger than the potential $v$  the  Eqs  (\ref{e31}),(\ref{e32}) can be
 solved for a very large number of values of $e_1$ and $e_2$ (dependent upon the
 values of $L$ and $c$).  But 
  when $e_1$  or both $e_1$ and $e_2$   become very
 large these two  equations 
  have no solution for any value of $L$ and $c$.  
 \bigskip \noindent 
\protect \section{The $v>e$ case  \label{sec4}} \smallskip 
 We discuss, now, the  $v>e$ case.    The matrix equations (\ref{e4}),(\ref{e5}) 
  may also be used for the  $v>e$ case but the middle matrix 
 $T$ at  the right hand  side of Eq (\ref{e5}) has to be written as 
 \begin{eqnarray}  && T_{11} =
 \cosh(\frac
{aq}{N})+\frac{\grave \xi}{2}\sinh(\frac{aq}{N}), \; \; \; \;  T_{12}
=\frac{\grave \eta}{2}
\sinh(\frac{aq}{N}) \label{e33} \\ && T_{21} = -\frac{\grave \eta}{2}
\sinh(\frac{aq}{N}), \; \; \; \;  T_{22} =
 \cosh(\frac
{aq}{N})-\frac{\grave \xi}{2}\sinh(\frac{aq}{N}) \nonumber  \end{eqnarray}
  $k$ is
  $\sqrt{\frac{2me}{\hbar^2}}$, 
  $q$ is $\sqrt{\frac{2m(v-e)}{\hbar^2}}$,    
 and $\grave \xi$ and  $\grave \eta$ are given by  
$\grave \xi=-i\eta=\frac{q}{ik}+\frac{ik}{q}$, \quad  
 $\grave  \eta=-i\xi=\frac{q}{ik}-\frac{ik}{q}$,  
 where $\eta$ and $\xi$ are from Eq (\ref{e7}).
 We can continue through the same steps as those of the 
 $e>v$ case and find  that the equivalent of  the $f$,  $d$, and $\phi$  from
 Eqs 
 (\ref{e12}),(\ref{e13})  are
\begin{equation} \label{e34} \grave f=kb-\frac{aq\eta}{2},  \; \; \; 
\grave d=\frac{aq\xi}{2}, \; \; \; \grave \phi^2=(\grave f \sigma_3-i\grave d\sigma_2)^2=
\grave f^2-\grave d^2
\end{equation} and the corresponding equations to (\ref{e14})-(\ref{e15})  are 
\begin{equation} \label{e35} e^{i((kb-\frac{aq\eta}{2})\sigma_3-
\frac{iaq\xi}{2}\sigma_2)}=\cos(\sqrt{\grave f^2-\grave d^2})+
\frac{i(\grave f\sigma_3-i\grave d\sigma_2)}
{\sqrt{\grave f^2-\grave d^2}}\sin(\sqrt{\grave f^2-\grave d^2}) \end{equation} 
 \begin{equation} \label{e36} \left[ \begin{array}{c} A_{2N+1} \\ B_{2N+1}
\end{array} \right]=\left[ \begin{array}{c c}
e^{-iz}(\cos{\grave \phi}+\frac{i\grave f\sin(\grave \phi)}{\grave \phi}) 
&-ie^{-iz}\frac{\grave d\sin(\grave \phi)}{\grave \phi} 
\\ ie^{iz}\frac{\grave d\sin(\grave \phi)}{\grave \phi}&
e^{iz}(\cos{\grave \phi}-
i\frac{\grave f\sin(\grave \phi)}{\grave \phi})
\end{array} \right]\left[ \begin{array}{c} A_{0} \\ B_{0}
\end{array} \right] \end{equation}
From the last two equations we can find the transmission probability  for the
$v>e$ case (in an analogous way to the $e>v$ case)
\begin{equation} \label{e37} |\frac{A_{2N+1}}{A_0}|^2=
|\frac{1}{e^{iz}(\cos(\grave \phi)-
\frac{if\sin(\grave \phi)}{\grave \phi}})|^2= |\frac{e^{i(\grave \kappa-z)}}
{\sqrt{\cos^2(\grave \phi)+
\frac{\grave f^2(\sin^2(\grave \phi)}{\grave \phi^2}}}|^2 
=\frac{1}{1+\frac{\grave d^2(\sin^2(\grave \phi)}
{\grave \phi^2}} \end{equation}
The $e^{i\grave \kappa}$ is the same as the $e^{i\kappa}$ from the previous section (see 
the inline equation prior to Eq (\ref{e18}))
except that we substitute  from Eq (\ref{e34}). 
Here, as for the $e>v$ case, the last expression  reduces, when $b=0$,  
to the known transmission probability \cite{Merzbacher} for the one barrier 
located at  the   same place  and exposed to the same wave function.  
\begin{equation} |\frac{A}{A_0}|^2= 
\frac{1}{\cosh^2(aq)+\frac{\eta^2(\sinh^2(aq))}{4}} \label{e38} \end{equation} 
As for the $e>v$ case the presence of a finite $b$ yields a new possibility for
the transmission probability to be 1 even when $v>\!>e$. Figure 10 is a three 
dimensional surface of the transmission  probability from
 Eq (\ref{e37}) as a function of the energy $e$ and $c$.  The range of $e$ is 
$150 \le e \le 192$, and that of $c$ is  $0.01 \le c \le 5$ which is the same
range  as in figure 6. The potential $v$ is 200. As for the $e>v$ case the
transmission probability tends to unity as $c$ increases. 
\par
We can continue, in a parallel way to the $e>v$ case,  and  find 
 the scattering cross section  from the corresponding $\grave S$-matrix which is 
 found from Eq (\ref{e36}) (We denote the matrix in this equation by $\grave Q$) to be    
  $\grave S= \frac{1}{\grave Q_{22}} 
 \left[ \begin{array}{c c} 
  1&\grave Q_{12} \\ \grave Q_{21} &1 
\end{array} \right]$.  From the Eq 
$\det(\grave S-\grave \lambda I)=0$ (see the equivalent discussion for the $e>v$ case) 
  we can determine the corresponding
eigenvalues $\grave \lambda_{\pm}$ which are the same as those found in Eq 
(\ref{e22}) except for the different $\grave f$,  $\grave d$ and $\grave \phi$. From 
these 
$\grave \lambda_{\pm}$ we can write  equations corresponding  to 
(\ref{e23}),(\ref{e24}),   and from these equations 
the cross section $\grave \sigma_{\pm}$ may be obtained in an equivalent way to that of 
the $e>v$ case. It is found that also here  the period of 
$\grave \sigma_{\pm}$  becomes larger as $e$ increases. 
 As for the $e>v$ case,  if we release the condition of
 a constant $L=a+b$,  we find that  the dependence of the 
cross sections $\grave \sigma_{\pm}$, as  functions of the energy $e$,   upon 
the total interval $b$ is non-trivial only for a
specific finite range which depends upon $a$. For example, for $v=140$,  $e$ 
in the range $1 \le e \le 120$, and $a=10$ 
the cross section   $\grave \sigma_+$, as a function of $e$,  changes 
with $b$ only in  the range $ 0 < b \le 8.7a$.   For any other value of  
$b>8.7a$ we obtain $\grave \sigma_+=0$. The range of dependence upon $b$ becomes
smaller as $a$ grows.  for example, when $a$ grows from the former value of 10 
to
15 the former range of $b$ becomes smaller by almost a factor of 3, so that the
new range in which  $\grave \sigma_+$, as a function of $e$,    changes with 
  $b$ is $ 0 < b \le 3.13a$. That is, for any other
 value of  $b>3.13a$ we obtain $\grave \sigma_+=0$.  \par 
 We note that although 
 the periods of $\grave \sigma_{\pm}$ become larger as the
energy $e$ grows,  the rate of  growth is  smaller compared to that of the
$e>v$ case, and  as for the later case, the total width $a$ and the rate
of growth of the periods of $\grave \sigma_{\pm}$ are inversely proportional. That is,
as $a$ grows the growth rate of these periods becomes small.   
\par 
We discuss now the issue of resonances for the $v>e$ case. We may use, for that
matter,  the equation corresponding to Eq  (\ref{e29}) of the $e>v$ case, except that
we substitute  the $\grave f$, $\grave d$ and $\grave \phi$ from Eq 
(\ref{e34}) and also the $q$ of the $v>e$ case. We obtain the 
following equation
\begin{eqnarray}  
&&\sin(\sqrt{\frac{(a+ac+cL)}{(1+c)^2}(e(a+ac+cL)-
 av(1+c))})= \label{e39}  \\ &&=
 \pm \frac{2i}{av(1+c)}\sqrt{e(a+ac+cL)(e(a+ac+cL)-va(1+c))} \nonumber 
 \end{eqnarray}
We now differentiate between two cases: The first is when 
$va(1+c)<e(a+ac+cL)$ in which case 
the last equation is identical to Eq (\ref{e30}), and so we can use  the two simultaneous equations
 (\ref{e31}),(\ref{e32}) in order to find the real and imaginary parts
 $e_1$ and  $e_2$ of the energy $e$. The essential difference between 
 Eq (\ref{e30}) and  Eq (\ref{e39}) (when 
 $va(1+c)<e(a+ac+cL)$) is that in  Eq (\ref{e30})
  we have $e>v$, whereas here  $v>e$. 
  As for the $e>v$ case,  the allowed ranges of $c$ depend upon the 
 values of $L$ such that as $L$ increases these ranges grow.    \par
We can prove, in an analogous manner to the $e>v$ case (see Appendix B), that the energies 
that may be
considered as poles of the cross sections $\grave \sigma_{\pm}$ can not assume 
very large values (although here these energies have to satisfy the condition 
$v>e$). \par
The second case is
 when we have  in Eq (\ref{e39})  $va(1+c)>e(a+ac+cL)$, 
 in which we obtain  
 \begin{eqnarray} 
  &&  i\sinh(
 \sqrt{\frac{(a+ac+cL)}{(1+c)^2}(av(1+c)-e(a+ac+cL)
 )}) =  \label{e40}  \\
 &&=\mp  \frac{2}{av(1+c)}\sqrt{e(a+ac+cL)(va(1+c)-e(a+ac+cL))} \nonumber  
 \end{eqnarray} This  equation can be solved  only  for complex energies   
 $e=e_1+ie_2$.  As in the $e>v$ case we eliminate the square roots and the
 complex character  
 from both sides of Eq (\ref{e40}) by using the deMoivre theorem from Eq (\ref{eA1}) in  
 Appendix A  and 
 the following hyperbolic sine addition formula \cite{Schaum} 
  $\sinh(x+iy)=\sinh(x)\cos(y)+i\cosh(x)\sin(y)$.  
 Thus,     
 comparing separately the reals and imaginaries we obtain the following two simultaneous 
 equations from which we try to find the real and imaginary parts of the energy
  \begin{equation} \label{e41} 
-\cosh(r_1^{\frac{1}{2}}\cos(\frac{\phi_1+2\pi k}{2}))
 \sin(r_1^{\frac{1}{2}}\sin(\frac{\phi_1+2\pi k}{2}))=
 \mp r_2^{\frac{1}{2}}\cos(\frac{\phi_2+2\pi k}{2}) \end{equation} 
 \begin{equation} \label{e42} \sinh(r_1^{\frac{1}{2}}
 \cos(\frac{\phi_1+2\pi k}{2}))
 \cos(r_1^{\frac{1}{2}}\sin(\frac{\phi_1+2\pi k}{2}))=
\mp  r_2^{\frac{1}{2}}\sin(\frac{\phi_2+2\pi k}{2}),  \end{equation}
where $k=0, 1$, 
 and $\grave r_1$, $\cos (\grave \phi_1)$, $\sin (\grave \phi_1)$, 
 $\grave r_2$, $\cos (\grave \phi_2)$  and $\sin (\grave \phi_2)$ are given respectively 
 by 
 \begin{eqnarray} && \grave r_1=r_1, \; \; \;  
  \sin(\grave \phi_1)=-\sin(\phi_1), \; \; \;   
 \cos(\grave \phi_1)=-\cos(\phi_1) \label{e43} \\ 
&& \grave r_2=r_2, \; \; \; 
 \sin(\grave \phi_2)=\cos(\phi_2), \; \; \;  
 \cos(\grave \phi_2)=-\sin(\phi_2)  \nonumber \end{eqnarray}
The variables $r_1$, $\cos(\phi_1)$,   $\sin(\phi_1)$,  $r_2$, $\cos(\phi_2)$ 
and $\sin(\phi_2)$ are those of the $e>v$ case and are given by equations
(\ref{eA3})-(\ref{eA8}) in Appendix A. 
 It has been turned out,  numerically,  that there is no solutions to the  two
simultaneous equations (\ref{e41}),(\ref{e42}) that satisfy  the condition of 
$va(1+c)>e(a+ac+cL)$. \par
We note that when we have released the
condition of constant $L$ we find (see the discussion on the cross-section 
$\grave \sigma_{\pm}$ before  Eq (\ref{e39})) 
that the 
cross sections $\grave \sigma_{\pm}$ become and remain zero for all values of 
$b$ that exceed some limiting value  
that depends upon the values of $a$. 
We have, also,  found that these limiting values of
$b$ become smaller as $a$ becomes larger.  That is, in these cases the cross 
sections $\grave \sigma_{\pm}$ certainly have no poles. Now, when $a$ becomes
large the probability that the difference $va(1+c)-e(a+ac+cL)$ will be positive
increases, and in this case,  as we have just found,  the cross sections 
$\grave \sigma_{\pm}$ have no poles in accordance with our discussion here. \par 
Summarizing the subject of poles  in the last two sections we see 
that for the  $|e|>v$ case we find a very large number of different poles in 
large ranges of $c$, where the extent of these  ranges depends upon the values 
of $L$.  When 
$v>|e|$  the existence of poles depends upon the difference  
$va(1+c)-|e|(a+ac+cl)$. That is, if this difference is negative then poles are
found to the scattering cross sections $\grave \sigma_{\pm}$, although in
smaller ranges of $c$ compared to the corresponding ranges of the $|e|>v$ case.
When the difference  $va(1+c)-|e|(a+ac+cL)$ is positive  no pole of 
$\grave \sigma_{\pm}$ is found.  \bigskip \noindent 
\protect \section{concluding remarks  \label{sec5}} \smallskip 
We have discussed the properties of a large number $N$ of one-dimensional potential
barriers arranged in a finite region  of the $x$ axis.  
We use both 
a $4N \times 4N$ matrix method for finite $N$ and the transfer matrix 
method for the infinite array of these potential barriers along the finite
region.   We have
discussed both cases of $e>v$ and $v>e$ and for both cases we found 
that the ratio of the total intervals between these potential barriers to 
their total width  is an important parameter that determines  the
properties of the above mentioned variables. For example, when this ratio
increases the transmission coefficient, for both cases
 of $e>v$ and $v>e$, of the passing 
plane wave or wave packet tends to the unity value even when the initial
energies of these waves are very small. A similar effect was found \cite{Bar1} 
in a classical diffusion system with a high density of imperfect traps for which
the survival probability \cite{Smol} of classical particles passing through it tends
to unity when the interval between the traps increases. Another system that was
found \cite{Peres,Bar2} to demonstrate the same behaviour is the array of identical
optical analyzers, such as Nicol prism \cite{Born}, so that when the number of them,
along a finite interval, becomes very large a beam of light passes through them
with the same initial polarization and intensity it had before the passage. \par   
 We have shown in this paper that a potential constructed of a large
number of identical barriers can induce the  type of  behaviour as observed
in the neighbourhood of tunneling barriers \cite{Schieve}, interpreted in these
references as chaotic-like. One may consider, 
as for the parallel drawn \cite{Schieve} between an unstable 
fixed point of
the classical problem and  quantum chaotic-like behaviour for the single
barrier tunneling problem, a classical analog to the problem studied here. The
repetitive potential in the bounded region, approached from above (for $e>v$),  
appears  as an accumulation of unstable fixed points. The single wide barrier, 
on the other hand, is quasi-stable when approached from above; it is only the
tunneling configuration in this case that has a strong analogy to the effect of
a separatrix. For $c$ large, when the potential barriers are relatively well
separated, we see an apparent chaotic-like effect most strongly through the Wigner type
level distribution, and when $c$ is small the distribution moves toward Poisson
type.

\bigskip \noindent \protect \section*{\bf Acknowledgement }
\bigskip  \noindent   We wish to thank  D. Pearson and W. Amrein 
 for discussions at an early stage of 
this work

\appendix
\section{}
  Eqs (\ref{e31})-(\ref{e32}) are obtained after  eliminating 
  the square roots and the complex nature from both sides of Eq (\ref{e30}). We do this  
  by 
  using the following two trigonometric relations  \cite{Schaum}.   \begin{equation} 
  \label{eA1}
  (r\cos(\phi) 
 +ir\sin(\phi))^{\frac{1}{n}}=r^{\frac{1}{n}}(\cos(\frac{\phi+2\pi k}{n})+
 i\sin(\frac{\phi+2\pi k}{n})),  \end{equation}  where $n$ is any positive integer and 
 $k=0, 1, 2 \ldots n-1$. 
 \begin{equation} \label{eA2} \sin(a \pm ib)=\sin(a)\cos(ib) \pm \cos(a)\sin(ib)=
 \sin(a)\cosh(b) \pm i\cos(a)\sinh(b), \end{equation}
  Comparing separately the real 
 and the imaginary parts  of both sides  we 
 obtain the  two simultaneous 
 Eqs (\ref{e31}),(\ref{e32}) from which we can determine the  components 
 $e_1$ and $e_2$ of the energies $e$ that satisfy Eq (\ref{e30}).  The six variables of 
Eqs (\ref{e31}),(\ref{e32}) that depends upon the     cordinates $(r,\phi)$ are given by 
 \begin{equation} r_1=\sqrt{\frac{(a+ac+cL)^2}{(1+c)^4}((e_1(a+ac+cL)-
 va(1+c))^2+(e_2(a+ac+cL))^2)} \label{eA3} \end{equation}  
 \begin{equation} \label{eA4} \cos(\phi_1)= \frac{e_1(a+ac+cL)-
 av(1+c)}{\sqrt{(e_1(a+ac+cL)-av(1+c))^2+(e_2(a+ac+cL))^2}}
  \end{equation} \begin{equation} \label{eA5} \sin(\phi_1)=
  \frac{e_2(a+ac+cL)}{\sqrt{(e_1(a+ac+cL)-av(1+c))^2+(e_2(a+ac+cL))^2}} 
 \end{equation}  \begin{eqnarray} 
&& r_2=  \label{eA6} \\  &&=
\sqrt{\frac{16(a+ac+cL)^2}{a^4v^4(1+c)^4}((e_2va(1+c)-2e_1e_2(a+ac+cL))^2+
((a+ac+cL)(e_1^2-e_2^2)-
e_1va(1+c))^2)} \nonumber \end{eqnarray} \begin{equation} \label{eA7} 
 \cos(\phi_2) =
  \frac{e_2va(1+c)-2e_1e_2(a+ac+cL)}
 {\sqrt{(e_2va(1+c)-2e_1e_2(a+ac+cL))^2+((a+ac+cL)(e_1^2-e_2^2)-
e_1va(1+c))^2}}  \end{equation} \begin{equation} 
 \label{eA8} \sin(\phi_2)= \frac{(a+ac+cL)(e_1^2-e_2^2)-e_1va(1+c)}
 {\sqrt{(e_2va(1+c)-2e_1e_2(a+ac+cL))^2+((a+ac+cL)(e_1^2-e_2^2)-
e_1va(1+c))^2}}   \end{equation}

\section{} 
We show that there is no solution to Eqs  (\ref{e31})-(\ref{e32}) for very large 
values of $e_1$ or of both $e_1$ and $e_2$. In the first case we have $e_1>\!>e_2$ 
and we obtain from equations 
 (\ref{eA3})-(\ref{eA8}) In Appendix A 
 \begin{eqnarray} && \sin(\phi_1) \approx \cos(\phi_2) \approx 0, \; \; \;
 \cos(\phi_1) \approx \sin(\phi_2) \approx 1, \label{eB1} \\ 
 && r_1 \approx \frac{e_1(a+ac+cL)^2}{(1+c)^2}, \; \;
 \; r_2 \approx \frac{4e_1^2(a+ac+cL)^2}{a^2v^2(1+c)^2} \nonumber \end{eqnarray} Using these
 approximations we can write the two simultaneous equations 
 (\ref{e31}),(\ref{e32}) for the $k=0$ as (we note that the following two  
 equations do not change their forms   if this $k$ 
   assumes its second value of $k=1$). 
\begin{equation} \label{eB2} 
 \sin(r_1^{\frac{1}{2}}
  \cos(\pi k_1))\cosh(r_1^{\frac{1}{2}}
  \sin(\frac{\pi k_1}{2}))=\pm r^{\frac{1}{2}}_2\cos(\frac{\pi}{4}+
  \frac{\pi k_1}{2})
  \end{equation} \begin{equation}  \label{eB3} \cos(r_1^{\frac{1}{2}}
  \cos(\pi k_1))\sinh(r_1^{\frac{1}{2}}
  \sin(\frac{\pi k_1}{2}))= \pm r^{\frac{1}{2}}_2\sin(\frac{\pi}{4}
  +\pi k_1)
  \end{equation} 
  It can be seen that  Eq (\ref{eB3}) is not satisfied  
 for $k_1=0$ or an 
even $k_1$. We, now, show that these equations  are not
satisfied for any uneven $k_1$ either. For these $k_1$ these two equations can be
written as \begin{equation} \label{eB4} 
 \sin(r_1^{\frac{1}{2}})\cosh(r_1^{\frac{1}{2}})=
 \pm r^{\frac{1}{2}}_2\sin(\frac{\pi}{4})
  \end{equation} \begin{equation}  \label{eB5} \cos(r_1^{\frac{1}{2}})
  \sinh(r_1^{\frac{1}{2}})=\pm r^{\frac{1}{2}}_2\cos(\frac{\pi}{4})
  \end{equation}   
Squaring the two sides of both equations  we realize that their
right sides are the same.   So equating the left sides 
 we obtain \begin{equation} \label{eB6} 
\tan^2(r_1^{\frac{1}{2}})=\tanh^2(r_1^{\frac{1}{2}}) \end{equation} 
In order for the last equation to be valid the variable $r_1^{\frac{1}{2}}$ 
must be small, but it is given that $e_1$ is very large, so $r_1^{\frac{1}{2}}$ 
must also be very large (see  Eq (\ref{eB1}))
 and the
equation (\ref{eB6}) can not be satisfied. Thus, when 
 $e_1>\!>e_2$ we find no poles of the cross section from (\ref{e29}).
\par 
The same consequence is obtained also when both $e_1$ and $e_2$ are very large, 
so that we can write  $e_1\approx e_2$. In this case we obtain  
$$ \sin(\phi_1) \approx \cos(\phi_1) \approx \frac{1}{\sqrt{2}}, \; \; \; \;
 \cos(\phi_2) \approx -1, \; \; \; \;\sin(\phi_2) \approx 0, $$  
 and the corresponding two approximate simultaneous equations for  this 
 case are
 \begin{equation} \label{eB7} \sin(r_1^{\frac{1}{2}}\cos(\frac{\pi}{8}+\frac{\pi k_1}{4}))
 \cosh(r_1^{\frac{1}{2}}
  \sin(\frac{\pi}{8}+\frac{\pi k}{4}))=\pm r^{\frac{1}{2}}_2
  \cos(\frac{\pi+2\pi k_1}{2}) \end{equation}
    \begin{equation} \label{eB8}  \cos(r_1^{\frac{1}{2}}
  \cos(\frac{\pi}{8}+\frac{\pi k_1}{4}))\sinh(r_1^{\frac{1}{2}}
  \sin(\frac{\pi}{8}+\frac{\pi k_1}{4}))= \pm r^{\frac{1}{2}}_2
  \sin(\frac{\pi k_1}{2}) \end{equation} 
      It can be seen that the first equation can not be solved for any $k_1$ 
      (even or uneven).  Thus,  we see that very large energies can not 
      be  solutions of  (\ref{e31}),(\ref{e32}).   
 
 \bigskip \bibliographystyle{plain}

\begin{thebibliography}{99}
\bibitem{Remark1}     It can be shown that the matrix $T^N$ does not depend also
                       on the number of potential barriers $N$. That is, its
		       elements as given by the set (\ref{e3}) receive the same
		       values for each positive integer $N$.
\end{thebibliography}

\begin{figure}[hb]
\centerline{
\epsfxsize=3in 
\epsffile{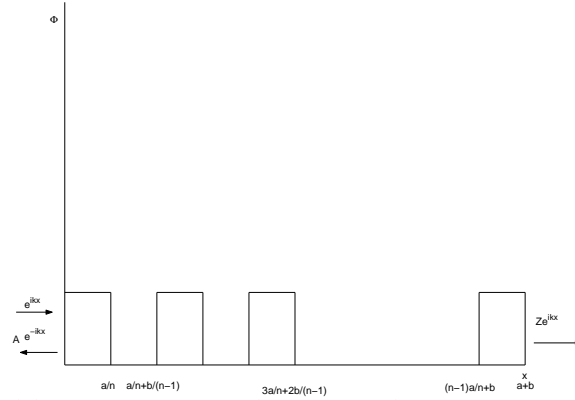}}
\caption[fig1]{ The $n$ potential barrier system. The approaching, transmitted
and reflected waves are shown at right and left.}
\end{figure}

\begin{figure}[hb]
\centerline{
\epsfxsize=3in 
\epsffile{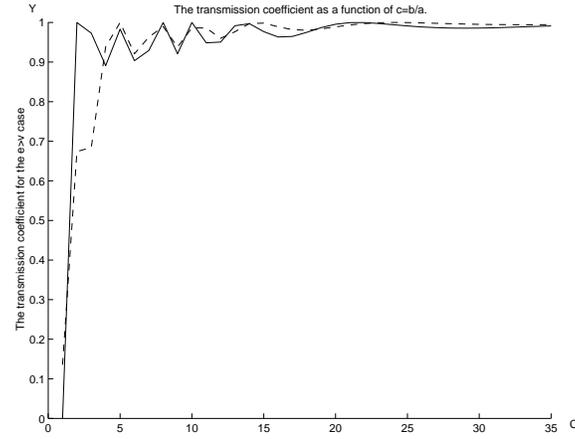}}
\caption[fig2]{ The continuous curve is the plot of the transmission
coefficient for $N=30$, and the dashed one is for $N=40$, both as
functions of $c$, and for the $e>v$ case.  Note that for the larger $N$ the
transmission coefficient tends to unity for smaller values of c. }
\end{figure}

 \begin{figure}[hb]
\centerline{
\epsfxsize=3in 
\epsffile{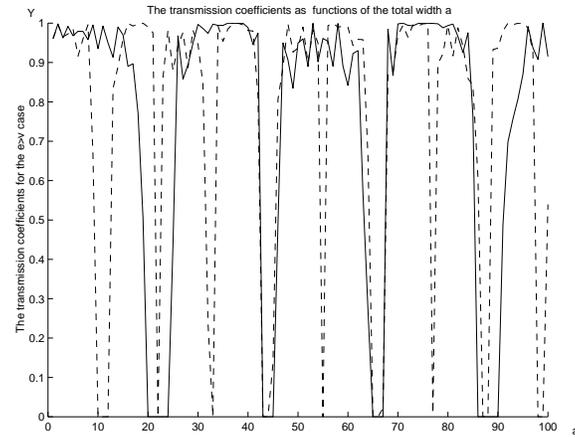}}
\caption[fig3]{ The dashed curve is the plot of the transmission
coefficient for $N=60$, and the continuous curve is for $N=120$, both as 
functions of $a$, and for the $e>v$ case. }
\end{figure}

\begin{figure}[hb]
\centerline{
\epsfxsize=3in 
\epsffile{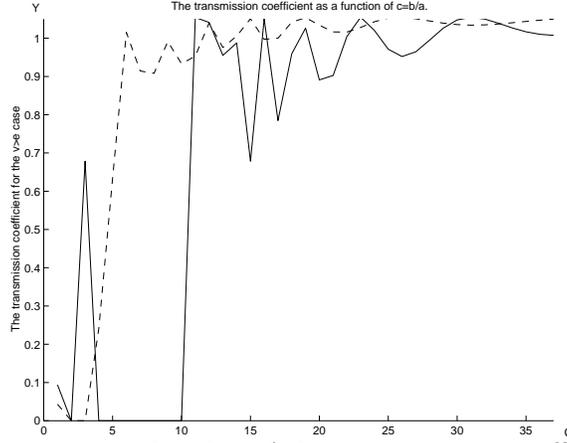}}
\caption[fig4]{ The continuous  curve is the plot of the transmission
coefficient for $N=30$, and the dashed one  for $N=50$, both as 
functions of $c$, and for the $v>e$ case. Note that as in figure 2 the
transmission coefficient for the larger $N$ tends to unity  at smaller
values of $c$.  }
\end{figure}

\begin{figure}[hb]
\centerline{
\epsfxsize=3in 
\epsffile{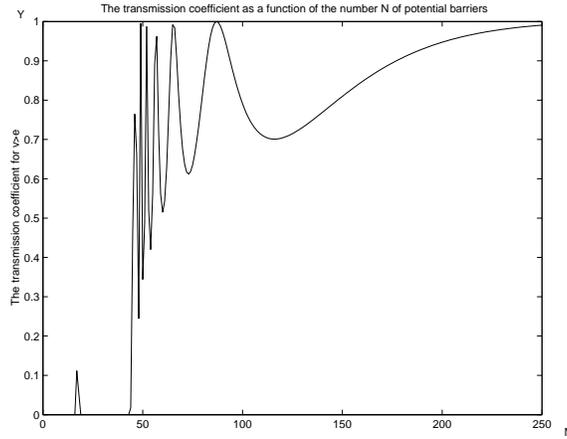}}
\caption[fegf5]{The transmission coefficient as a function of $N$ for the $v>e$ 
case. The total width a is 8, the potential $v$ is 202, and the energy $e$ is 
200. Note that this coefficient tends to unity as $N$ increases. }
\end{figure}

\begin{figure}[hb]
\centerline{
\epsfxsize=3in
\epsffile{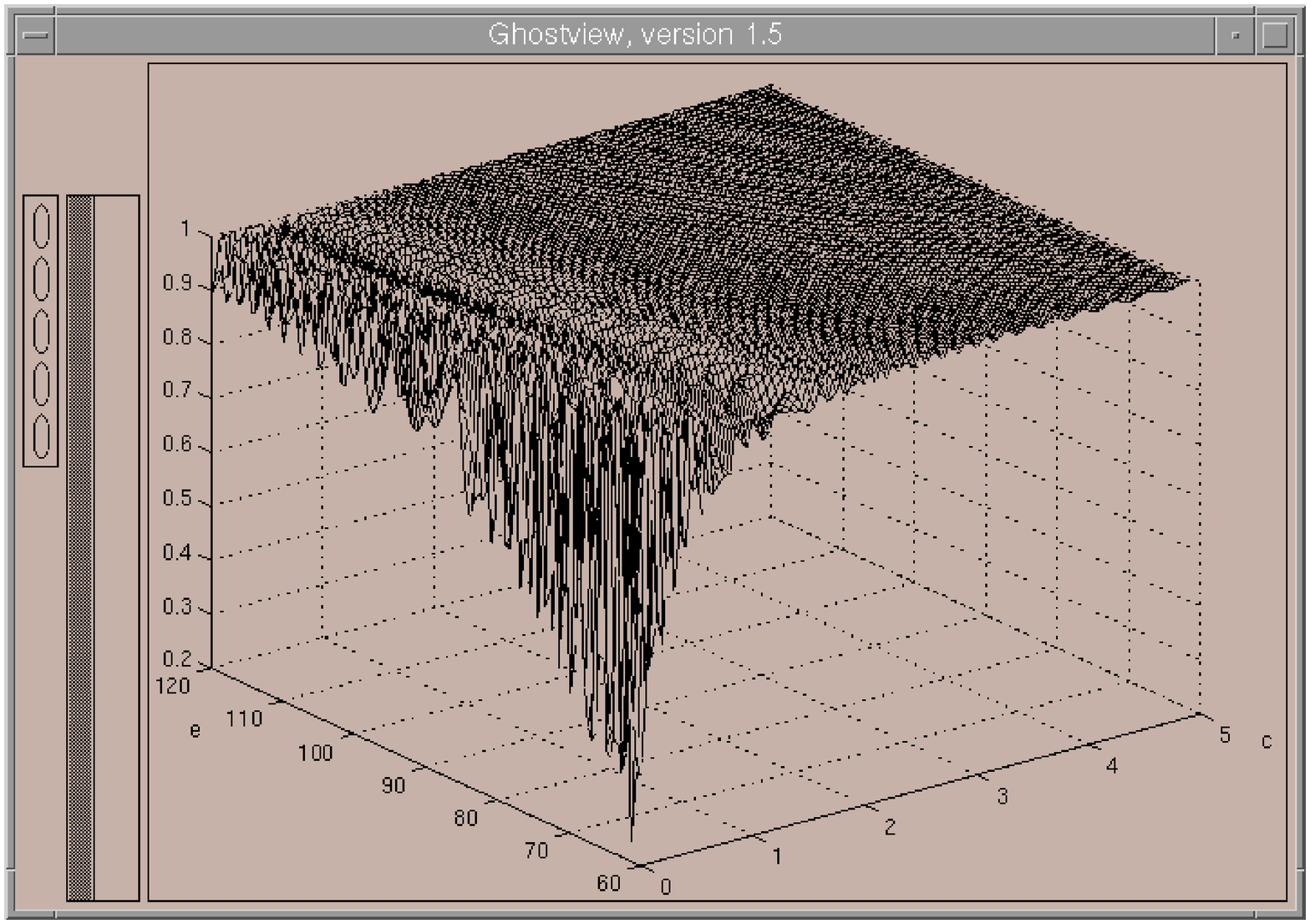}}
\caption[fegf6]{A three dimensional graph of the transmission probability from
Eq (\ref{e18}) for $e>v$ as a function of  $c$  and the energy $e$. This graph is 
for a total system length of
$L=70$,  $v=60$,  $c$ in the range 
$0.01 \le c \le 5$, and  $61 \le e \le 120$. Note that the
transmission probability as a function of $c$ tends to unity at a faster rate
than as a function of $e$. }
\end{figure}

\begin{figure}[hb]
\centerline{
\epsfxsize=2.5in
\epsffile{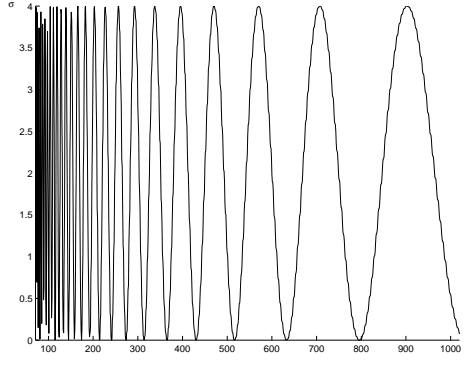}}
\caption[fegf3]{The graph of the cross section $\sigma_{+}$ from 
Eq (\ref{e25}) as a function of the energy $e$.
This graph is for a total length of $L=70$, $a=40$,  $v=70$, and for $e$ in the
range $71 \le e \le 1000$. Note that the period of the graph becomes larger as
the energy $e$ grows. }
\end{figure}

\begin{figure}[hb]
\centerline{
\epsfxsize=2.5in
\epsffile{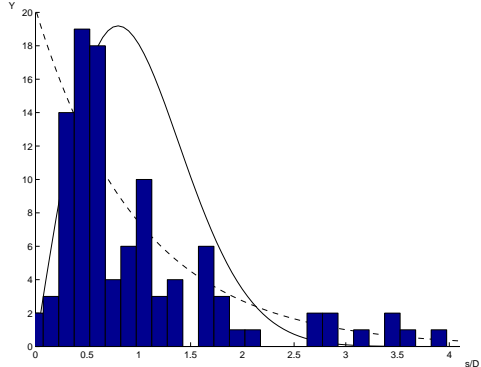}}
\caption[fegf8g ]{The histogram is the level spacing distribution as a function of
$s/D$ where $s$ is the spacing between neighbouring levels and $D$ is the mean
spacing. The histogram is constructed from  102 energy levels found in the range 
$1 \le e \le 600$, $v$ is 120, $C=90$, $L=20$, and $c=19$. The dashed curve
is the Poisson distribution and the continuous graph is the Chaotic Wigner one. }
\end{figure}
 
\begin{figure}[hb]
\centerline{
\epsfxsize=2in
\epsffile{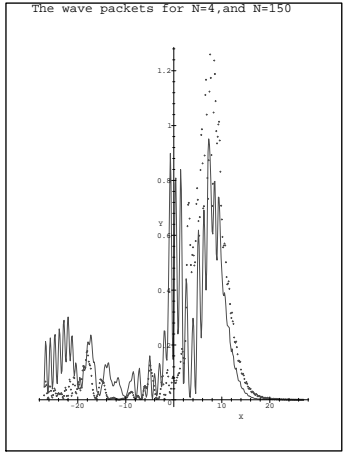}}
\caption[fegf9]{The continuous curve is the form of the wave packet  
for $n=150$, and the point-type curve is the form for  $n=4$. Both curves are  
at time $t=5.8$ in computer units.  }
\end{figure}

\begin{figure}[hb]
\centerline{
\epsfxsize=3in
\epsffile{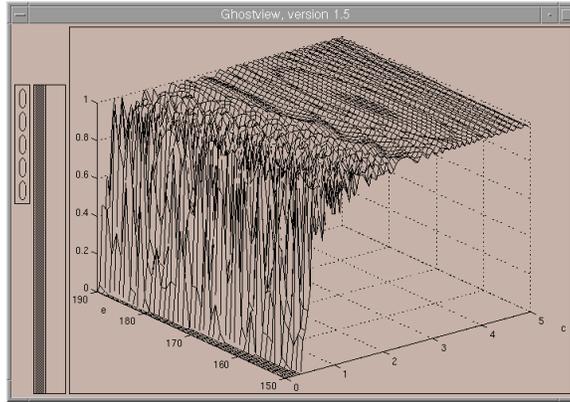}}
\caption[fegf10]{ A three dimensional graph of the transmission probability from
Eq (\ref{e37}) for $v>e$ as a function of $c$  and the energy $e$. This graph is
for $v=200$, $L=70$,  $c$ in the range 
$0.01 \le c \le 5$, and  $150 \le e \le 195$,   
We see from the figure that as the energy $c$ grows above the value 0.2 the transmission 
probability jumps  to 1. } 
\end{figure}
 
\end{document}